**Broad presence of ferromagnetism in bees and relationship to phylogeny, natural history, and sociality**

Laura Russo[1*], Caleb Allen[2], Cameron S. Jorgensen[2], Lizabeth Quigley[2], C. Charlotte Buchanan[2], Michael Winklhofer[4, 5], Seán G. Brady[6], Laurence Packer,[3] Anne Murray[1**], Dustin A. Gilbert[2†**]

*Corresponding author: lrusso@utk.edu

†Corresponding author: dagilbert@utk.edu

**These authors jointly supervised this work

[1]Department of Ecology and Evolutionary Biology, University of Tennessee, Knoxville, Tennessee, USA

[2] Department of Materials Science and Engineering, University of Tennessee, Knoxville, Tennessee, USA

[3]Department of Biology, York University, Toronto, Ontario, Canada M3J 1P3

[4]Institute of Biology and Environmental Sciences, University of Oldenburg, Oldenburg, Germany

[5]Research Center Neurosensory Science, Carl von Ossietzky University of Oldenburg, Oldenburg, Germany

[6]Department of Entomology, National Museum of Natural History, Smithsonian Institution, Washington DC, USA

**Abstract**

Scientists have long been fascinated by magnetoreception, the innate capacity of many animals to sense and use the Earth's magnetic field for navigation. In eusocial insects like honey bees, magnetoreception has been linked to communication and foraging. However, little is known about magnetoreception's phylogenetic patterns and relationship to species traits and natural history. Here, we demonstrate that putative magnetoreception based on ferromagnetic particles is widespread across a diversity of bee species (72 out of 96 species tested), with no phylogenetic signal. We also detected such putative magnetoreception in non-bee outgroups, suggesting this magnetic capacity predates the evolution of the Anthophila. While magnetic signals were found across a diversity of life history traits, the strength of the magnetic signal varied within and between species, and increased with body size and social behavior.

## Introduction

The innate biological capacity of some organisms to orient themselves across large spatial areas has captivated scientists for over a hundred years (*1–5*). While humans rely on tools like the compass for long-range navigation, many migratory animals possess an internal magnetic sense – magnetoreception – that allows them to detect and use the geomagnetic field. This ability has been demonstrated in animals such as salmon (*6*, *7*), lobsters (*8*, *9*), birds (*10*), turtles (*11*, *12*), and moths (*5*). Magnetoreception is thought to support navigation across long distances, particularly under conditions where visual cues (e.g., the sun, stars, landmarks, or polarized light) are absent or unreliable (*13*). Surprisingly, magnetoreception also appears in some non-migratory organisms, presumably for local orientation and short-range navigation. These include ants (*14*, *15*), honey bees (*2*, *16*), and wasps (*17*), among others (*4*).

Magnetoreception has been extensively studied in the honey bee, *Apis mellifera* (*2*, *16*, *18–24*) and has also been suggested to occur in four other eusocial bee species: two species of bumble bees *Bombus impatiens* (*25*) and *Bombus terrestris* (Nilsson, 2013; but see Taylor, 2024), and two stingless bees, *Tetragonisca angustula* (*28*) and *Schwarziana quadripunctata* (*29*); these five species represent the cumulative knowledge of magnetoreception in bees (Anthophila). Because the hymenopteran taxa in which magnetoreception has been most studied – bees and ants – are eusocial central-place foragers, it was hypothesized that this ability may have evolved not only to aid in navigating to and from food sources, but also to support the communication of resource locations to nest- or hive-mates (*30*). Such behavior has been documented – for example, honey bees use magnetic cues to orient their waggle dance (*31*, *32*), enabling them to accurately communicate navigational information. In this context, magnetoreception has been linked to social behavior in bees (*33*). However, this association has yet to be tested through comparisons with solitary bees or non-central-place foragers, such as cuckoo bees.

The bees (Anthophila) represent an excellent candidate taxon for investigating magnetoreception in non-migratory species, with particularly rich opportunities for testing the relationship between magnetoreception and sociality, because within the clade there exist species which exhibit the full range of sociality, including semi-social and cleptoparasitic (*34*, *35*). Therefore, if magnetoreception is an evolutionary byproduct of social behaviors in Hymenoptera, one would expect to see a significant association between magnetic properties and sociality. Moreover, one would assume that the trait of magnetoreception should affect fitness and therefore should exhibit a phylogenetic signal (*36*). However, reports of magnetism or magnetoreception in Anthophila are limited to a handful of species, insufficient to speculate on the phylogenetic origins of magnetoreception in the Hymenoptera, or whether any phylogenetic signal even exists. Furthermore, to our knowledge, no one has investigated the magnetic properties or magnetoreception of solitary bees, making the suggested correlation between sociality and magnetoreception one-sided.

In non-migratory, non-social species, magnetoreception has been shown to relate to orientation, for example in crustaceans (*8*) and beetles (*37*). It is possible that it allows animals to create a "magnetic map" allowing them to derive their spatial position in a way that is useful for foraging and/or dispersal behaviors (*38*). If this is true, then magnetic signal could also differ significantly between the sexes in bees, which exhibit different nutritional needs, and foraging and mating behaviors (*39*, *40*). Female bees actively feed on pollen as adults, and collect pollen to provision their offspring, while males consume much less pollen and primarily feed on nectar (*41*). Differences in diet may be important as the leading mechanistic theory of magnetoreception in bees requires the presence of ferrimagnetic nanoparticles, derived from iron found in pollen (*30*). However, all bees consume pollen in their larval stages, so the role of adult pollen consumption is not clear. Males may also disperse greater distances to seek reproductive partners (*42*), and if magnetoreception plays a role in dispersal, we may further expect a difference in the magnetic properties between the sexes. The nesting behavior of these insects may also relate to their magnetism. For example, while most bee species are ground-nesting, there are also cavity- and stem-nesting bees that may require different cues for navigating to and from their nests (*35*). Moreover, there are many bee species that do not build their own nests, but rather lay their eggs within the nests of other bee species. These cleptoparasitic bees likely have different navigational needs (*43*) as it has been shown in several species that females learn the location of host nests and return to them repeatedly (*35*), again emphasizing the need for high-quality short- to medium-range navigation. Finally, body size has been associated with the distance a given central-place forager can travel from their nest (*44*). Thus, we may expect that magnetic sense may correlate with body size.

There are two primary hypotheses for the mechanism underlying magnetoreception in insects (*45*), one based on magnetosensitive biochemical reactions involving a radical-pair (*46*), and one based on mechanical torques or forces produced by magnetic particles (*47*, *48*). Good candidates for the first mechanism are light-sensitive cryptochromes, such as Drosophila-type Cry1 (*49*), but their involvement in geomagnetic field sensing remains to be demonstrated beyond doubt (*49*, but see *50*). The best candidate for the second mechanism is sub-100 nm magnetite nanoparticles, which would not require light to work (*52*) and therefore could be located anywhere in the body, provided they are associated with sensory neurons. To narrow down the location of these magnetoreceptive tissues, a number of studies have focused on the detection of magnetic particles within the body. In bees, studies have demonstrated the presence of magnetic particles in a variety of body parts: for *Bombus* the wings (*53*, *54*), for *Schwarziana* and *A. mellifera* the antennae (*30*, *55*, *56*), and for *A. mellifera* the metasoma (i.e. abdomen, see supplemental material) (*24*, *30*, *55*). Other works have demonstrated ferromagnetism in the antennae of the ant species *Pachycondyla marginata and Solenopsis interrupta* (*57–59*) and the abdomen and thorax of termites (*60*). The appreciable variation in the location of magnetic tissue among the few known magnetoreceptive bees – only five species – is also surprising. Notably, ants and termites are also eusocial insects, reinforcing the perceived correlation between sociality and magnetoreception. Thus, the magnetite-mediated mechanism has been proposed for eusocial insects (*30*), but it is not clear whether the mechanism is the same in solitary insects.

We evaluated magnetic properties across a diverse array of 96 bee species from six families and 47 non-bee insects as an outgroup to compare magnetic properties and provide context for variation within the monophyletic bee clade. We propose that the presence of a ferromagnetic signal would indicate the existence of crystalline ferrimagnetic nanoparticles, likely magnetite ($Fe_3O_4$), and suggest a potential role in magnetoreception. It seems likely that a ferromagnetic signal points to magnetoreceptive tissues, though the possibility of contamination from environmental pollution is also considered and addressed in the discussion. Our goals were to determine whether: 1) magnetic properties were associated with sociality, sex, nesting behavior, or body size in bees, 2) magnetic properties exhibited a phylogenetic signal or a phylogenetic origin in the bees, 3) magnetic properties were restricted to a particular region or body part in bees, 4) magnetic properties were restricted to bees or could be found in other insects.

**Results**

We measured the magnetic response of 185 insect specimens, including 138 bee specimens representing 96 species or morphospecies in six families (Andrenidae, Apidae, Colletidae, Halictidae, Megachilidae, and Melittidae). We included bee species that exhibit a range of nesting and social behaviors, including ground-nesting, cavity-nesting, and stem-nesting bees exhibiting the full range of sociality (Table S1). We also included three species known for their nocturnal or crepuscular behaviors for which magnetism-based orientation might be especially important. Most (114, including four queens) of the bee specimens were females, but we also included 24 males to determine whether there was a difference in the presence of these magnetic nanoparticles between the sexes. Most bee species were represented by a single specimen, or one specimen for each sex, but we tested multiple female specimens for 13 of the species to check for variation among individuals. Beyond the bees, we tested 40 insects from outgroups in the broader Apocrita (including close bee relatives in the Crabronidae, Philanthidae, Pemphredonidae, Ampulicidae, and Sphecidae, and more distantly related Vespidae, Tiphiidae, Pompilidae, Scoliidae, Ichneumonidae, and Braconidae), as well as five flies (Conopidae, Syrphidae, Tachinidae) and two beetles (Cantharidae, Coccinellidae).

Magnetic measurements focused on room-temperature hysteresis loops, which capture the evolution of the magnetization as the applied magnetic field is swept from large positive to large negative values and back (Fig. 1A). The progression of magnetization between these extremes reveals the underlying magnetic ordering: most common materials respond linearly, indicating paramagnetism, diamagnetism, or antiferromagnetism. In contrast, sigmoidal magnetization curves (Fig. 1) are characteristic of ferro- or ferrimagnetism. At high fields, ferro- and ferrimagnetic materials exhibit saturation as their magnetic moments align with the applied field, whereas other magnetic orderings (e.g. antiferromagnetism) do not saturate. Conducting these measurements at room temperature reflects biologically relevant conditions and helps avoid spurious signals from naturally occurring iron-containing compounds, such as ferritin and hemosiderin, which have parasitic ferrimagnetism at lower temperatures (*61, 62*). Ferrimagnetism necessarily arises from superexchange coupled ferric/ferrous ions in a crystal lattice; isolated or dilute ferrous ions, such as the hemoglobin found in insect hemolymph, do not generate a sigmoidal magnetic response. Biogenic magnetic nanoparticles are generally considered to be ferrimagnetic iron oxide, which behave indistinguishable from conventional ferromagnets in the present measurements. Throughout the text we use the term ferrimagnet when discussing the material itself, and ferromagnet when describing its magnetic behavior or response.

From these specimens, we quantified the saturation magnetization ($M_S$) (magnetization in large fields, representing 'how magnetic' the insect is), remanence ($M_R$) ($M$ @ $H$=0, where $M$ is the magnetization and $H$ is the applied magnetic field, representing how resilient the magnetism is to retaining its orientation i.e. the 'magnetic memory' at zero field of the previously saturated state), and coercivity ($H_C$) ($H$ @ $M$=0, representing the magnetic field strength necessary to zero the magnetization and thus being a measure of magnetic stability) (Fig. 1A). Generally, the remanence is presented normalized to the saturation magnetization, forming the squareness ratio ($M_R/M_S$), which carries information about the magnetic domain state. Also, the magnetization values reported in emu g$^{-1}$ correspond to mass-specific magnetization, obtained by dividing the measured magnetic moment by specimen mass; the mass-specific saturation magnetization is identified by the variable $M_S/m$, distinct from the absolute magnetization $M_S$. These values tend to indicate nanoscale features of the magnetic particles: $M_S$ increases with particle size and number; $M_R$ and $H_C$ are often coupled and correspond to the magnetic anisotropy – a material property – and can indicate the quality of the crystalline ordering in the nanoparticles, but also tend to increase with particle size (up-to diameters of $\approx$80 nm) (*63*). The magnetization is expected to manifest from magnetic nanoparticles located within specialized magnetoreceptive tissue and to be composed of some form of iron oxide, likely magnetite ($Fe_3O_4$), but potentially also hematite (hexagonal $Fe_2O_3$), or maghemite (cubic $Fe_2O_3$). We included measurements of the whole body, supplemented with measurements on individual body parts (metasoma (abdomen), mesosoma (thorax, including wings and legs), and head (including antennae)) for species that were represented in our dataset by more than one specimen.

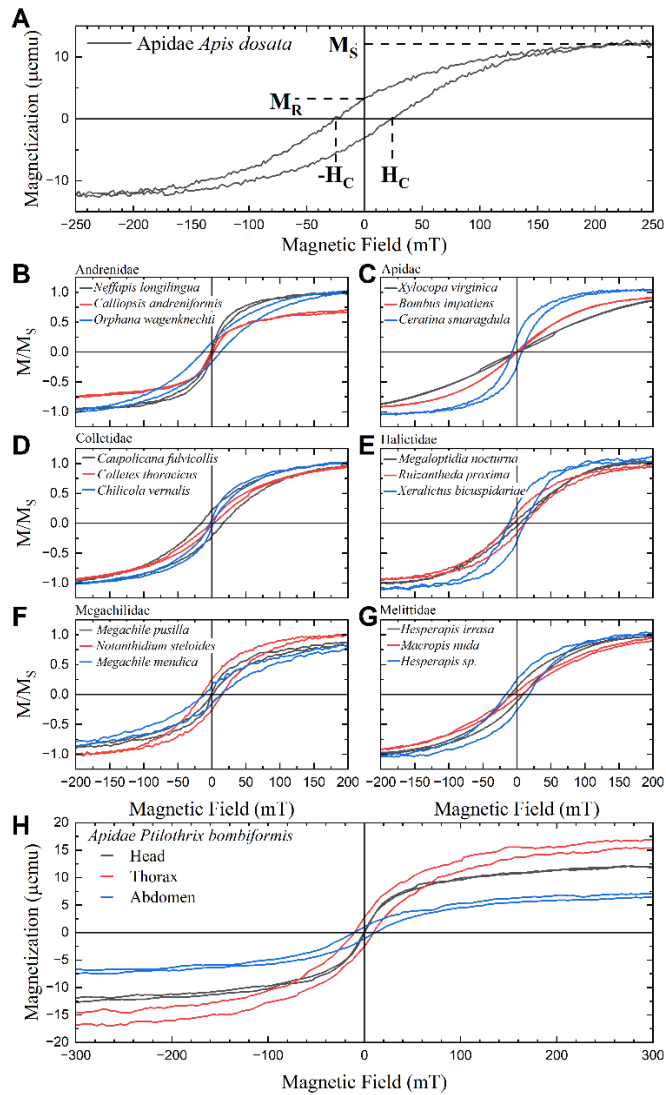

**Figure 1. Ferromagnetic responses are widespread across six bee families.** Magnetic hysteresis loop of Apidae *Epiclopus gayi* (A) illustrating the sigmoidal shape, remanent magnetization ($M_R$), saturation magnetization ($M_S$), and coercivity ($H_C$). Magnetic hysteresis loops from three representative individuals (indicated by color) measured in each of the six bee families (B: Andrenidae, C: Apidae, D: Colletidae, E: Halictidae, F: Megachilidae, G: Melittidae). These individuals were selected specifically to illustrate the appreciable variation in magnetic signals observed at the family level. Magnetic hysteresis loops measured for each of the three major body

parts of *Pilothrix bombiformis* (H), indicated by color (head = black, mesosoma = red, and metasoma = blue).

### *Whole insects*

Magnetic hysteresis loops were measured for each of 185 individual specimens, with at least 10 representatives from each bee family. Loops for ferromagnetic bees exhibited a sigmoidal response, indicative of ferromagnetic ordering, though the loops varied in overall shape and in quantitative metrics such as remanence and coercivity. As noted above, such differences likely correspond to variation in nanoparticle size, structure, or distribution. Notably, each family included species with open hysteresis loops and appreciable remanence ($M_R/M_S \approx 0.2$), suggesting that the particles behaved as stable magnets, with sufficient magnetic anisotropy to cause the magnetization to remain aligned with the previously saturated state. In these insects, the magnetic particles can be expected to experience a torque that attempts to rotate them physically into alignment with the external field, allowing north and south to be distinguished explicitly – thus enabling directional navigation. As the loops become narrower and remanence decreases, this torque diminishes and depends solely on the instantaneous field polarization (induced magnetization). In such cases, the magnetic field induces a net alignment of just the particle axes, resulting in a polarity-independent mechanism, making explicit north–south discrimination more difficult.

Insects that demonstrated a sigmoidal magnetization response, indicating ferromagnetism, but $H_C \approx 0$ and $M_R \approx 0$, were further tested at low temperatures (100 K) to probe for superparamagnetism, which occurs in ferro- or ferrimagnetic particles that are so small that thermal effects dominate their orientation, causing persistent re-orientation. For approximately spherical magnetite nanoparticles at room temperature, the transition from stable ferromagnetism to superparamagnetism typically occurs at particle diameters of $\approx 18$ nm. However, as discussed in (*64*), the precise threshold depends on the particle shape, aspect ratio, and magnetocrystalline anisotropy, which itself depends on crystal quality and defect density. Insects that showed a sigmoidal field response with non-zero $M_S$, $H_C$ and $M_R$ were classified as ferrimagnetic. Specimens with non-zero $M_S$ but $H_C \approx 0$ and $M_R \approx 0$ were re-measured at 100 K. If the saturation magnetization remained comparable upon cooling (typical changes $\approx 20\%$, as expected from the intrinsic temperature dependence following Bloch's $T^{3/2}$ law), these specimens were also classified as ferromagnetic. In contrast, specimens exhibiting a substantial increase in saturation magnetization upon cooling (typically $\approx 300\%$, following Curie–Langevin $1/T$ behavior), suggesting we passed through a blocking temperature, were classified as superparamagnetic. It is important to note that there is potential for contamination from ferromagnetic particles in the environment, however, as we discuss below, we expect this contamination to be relatively small (*65*).

At 300 K (room temperature), we detected a ferromagnetic response, i.e. a sigmoidal or hysteretic magnetization versus field progression, in 121 of the 138 bee specimens; 98 were females (including all four queens we tested), and 20 males (Table S2). Seventeen bee specimens (13 female, 4 male) exhibited a linear field response, with a positive (paramagnetic) slope, that we classified as nonmagnetic, and two of the female bees were superparamagnetic, exhibiting different responses at different temperatures, with discernable blocking temperatures. We observed variance within species for which we tested multiple specimens. In four of these cases, the female was nonmagnetic while the male was ferromagnetic (*Lasioglossum apocyni, Augochlorella aurata, Ceratina calcarata, C. mikmaqi*). For *Halictus confusus*, one female was ferromagnetic, while two females and a male were nonmagnetic, while for *Halictus ligatus/poeyi* and *C. strenua*, two male specimens were tested, one in each category (ferromagnetic, nonmagnetic) and the female was ferromagnetic. Similarly, for *Megachile mendica* we tested two specimens (both female) and found one ferromagnetic and one nonmagnetic. For *Apis mellifera*, we tested four workers, and one was superparamagnetic, while the other three and the male were all ferromagnetic; the queen *A. mellifera* was also ferromagnetic. For the non-bee insects, we detected ferromagnetism in 36 of 47 specimens, including at least one representative of each order we tested (Fig. 2A).

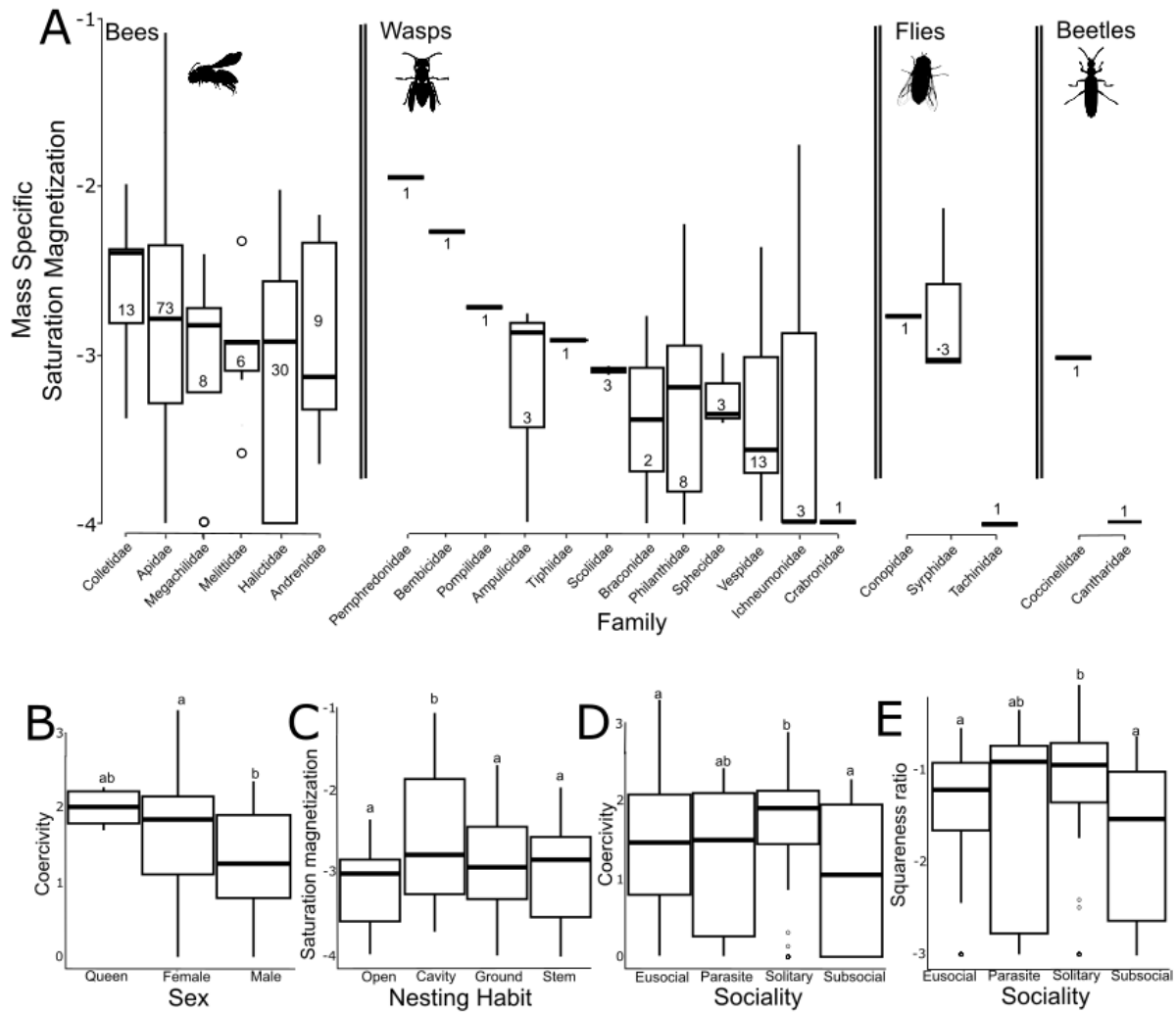

**Figure 2**. **Magnetic properties vary across insect taxa, and show significant relationships with sex, nesting habit, and sociality among bees.** Relationship between mass-specific saturation magnetization and each family of insects tested, included representatives of bees (far left), wasps, flies, and beetles (A). The number of specimens for each family is indicated by a number near each box. Relationship in bees between sex and coercivity (B). The letter above the box plots indicates significant differences between groups. Relationship in bees between nesting habit and mass-specific saturation magnetization (C). Relationship in bees between sociality and coercivity (D)

and between sociality and squareness ratio (E). Data are log$_{10}$-transformed for visualization and model fit (see Table S3 for precise values of transformation).

Interpreting the magnetization data for the bees, we first analyzed the magnetic coercivity. The coercivity is determined by the material, its quality, and the size of the nanoparticles; since the material and quality are expected to be the similar across the bees, the coercivity can be associated with the particle size, with larger particles having a larger $H_C$ (*63*). This analysis assumes that the particles remain in the single-domain regime, which occurs at ≈76 nm diameter. Across all bees, we found that female worker or solitary bees had a higher $H_C$ than male bees. Female worker or solitary bees did not differ significantly from the queens that were tested in three eusocial species (*Apis mellifera*, *Bombus griseocollis*, *Bombus pensylvanicus*) (Fig. 2B). Nesting habit was also significantly associated with the mass-specific saturation magnetization, $M_S/m$, such that the $M_S/m$ of cavity-nesting bees had significantly higher than ground-nesting bees, stem-nesting bees or bees that nest in the open (Fig. 2C). Nesting habit was not significantly associated with $H_C$ or $M_R/M_S$. Sociality was significantly associated with $M_R/M_S$ and $H_C$, but not $M_S$ (Fig. 2D, E). Eusocial bees had significantly higher $H_C$ than solitary bees, while solitary bees had a higher $M_R/M_S$ than eusocial or subsocial/semisocial bees. We found that the body size of the bees (intertegular distance, ITD) had a significant positive correlation with the $M_S$, $H_C$ and $M_R/M_S$ (Fig. S1, Table S3). This indicates that larger bees tended to have more and larger ferromagnetic nanoparticles compared to smaller bees.

Next, we use $M_S/m$ to identify species which possess substantive magnetic tissues and are therefore likely to exhibit magnetoreception. We identified a putative magnetoreception threshold using the $M_S$ and $M_S/m$ of *A. mellifera* (30 μemu, 0.7×10$^{-3}$ emu/g) (*66, 67*) and *S. quadripunctata* (3 μemu, 0.6×10⁻³ emu/g) (*56*, *68*). Since studies demonstrate that both of these species are magnetoreceptive, we use the lower limit of these values (3 μemu, 0.6×10⁻³emu/g) to set a threshold for 'potentially magnetoreceptive' (*16*, *20*, *23*). We consider both MS and MS/m because it is not known how magnetoreceptive tissues scale with insect size. These values are a conservative threshold based on the very few investigations of magnetoreceptive behavior in insects, the true threshold may be much lower. Using $M_S$ to define the threshold, we found that 144 of the 185 specimens were putatively magnetoreceptive, while 41 were not; of the bees specifically, 112 of 138 were putatively magnetoreceptive. The absolute magnetization, $M_S$, is not an ideal metric for comparison since tissue sizes generally scale with organism size, which could result in small bees with exceptional mass-specific magnetization, $M_S/m$, being below the threshold. Using $M_S/m$ to define the threshold, 120 specimens (94 bees) had putative magenetoreception, and 65 did not (Fig. 3A, B).

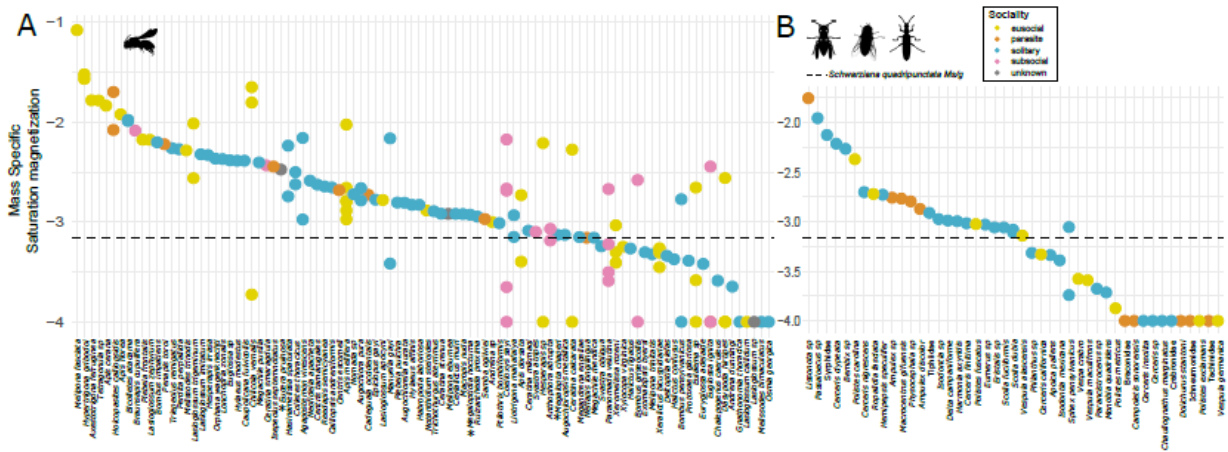

**Figure 3. Most bee and other insect species exceed the ferrimagnetic threshold set by a documented magnetoreceptive bee species.** Patterns of mass-specific saturation magnetization, $M_S/m$, across bees (A) and other insects, including wasps, flies, and beetles (B). Sociality is

indicated by color, where yellow is eusocial, orange is parasitic, blue is solitary, pink is subsocial or semisocial, and grey is unknown. The dotted black line indicates the $M_S/m$ for *S. quadripunctata* from the published literature. This line indicates thresholds for putative magnetoreception. Asterisks (*) indicate species that are nocturnal or crepuscular. The data are $\log_{10}$-transformed to improve visualization and model fit (see Table S3 for precise transformation values).

Among the bees, ferromagnetism and putative magnetoreception were widespread across all six families (*69*) (Fig. 4). We use the term ferromagnetism to describe insects which show the signature sigmoidal hysteresis loop but are below our threshold, while putative magnetoreception are ferromagnetic and are above the threshold. We did not find any evidence for a phylogenetic signal in $H_C$ , $M_S/m$ , or $M_R/M_S$. Based on our estimates of Blomberg's K statistic (*70*), we found significantly less phylogenetic signal than expected by Brownian motion for $H_C$ (K = 0.24), $M_R/M_S$ (K=0.52), and $M_S/m$ (K=0.10) (Fig. S2). We reran the linear models using *pgls*, but lambda (a phylogenetic scaling parameter) was equal to 0 and including the comparative data did not qualitatively change the outcome of the models.

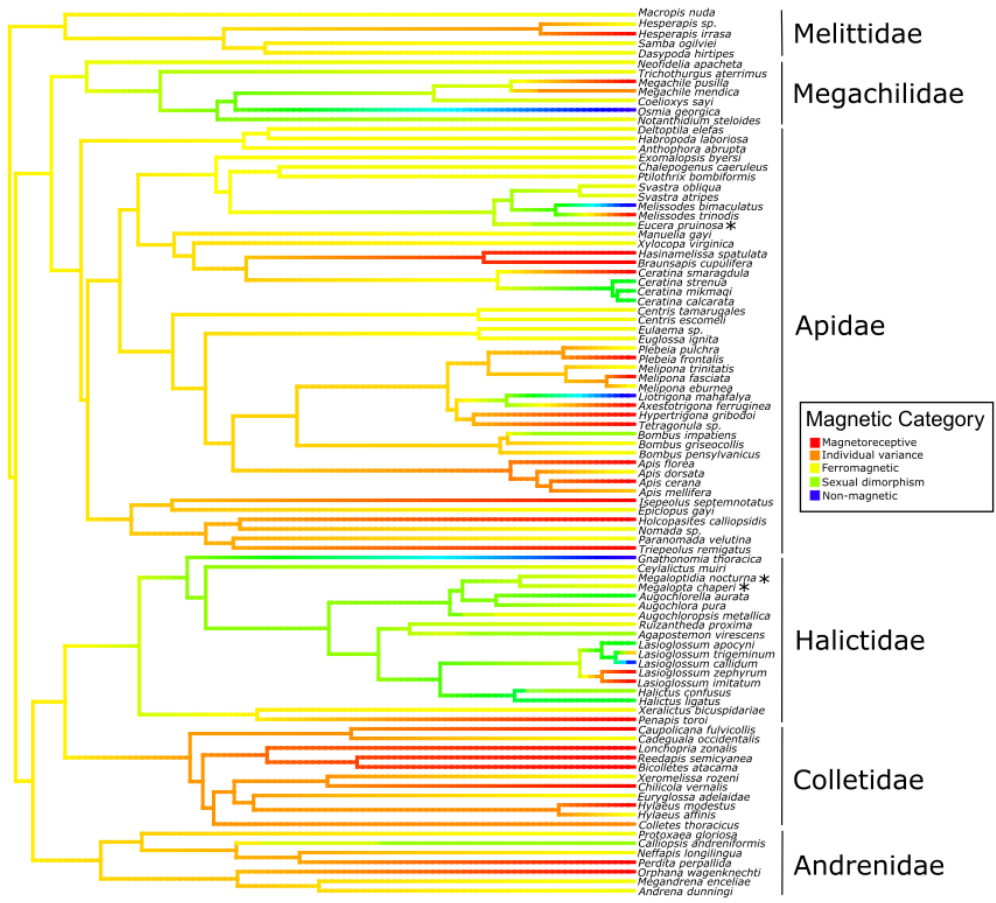

**Figure 4**. **Absence of phylogenetic signal in the magnetic response of bee species, suggesting the origin of magnetoreception predates the Anthophila.** A phylogenetic tree illustrating the bee species from six families we examined and their magnetic category (putative magnetoreceptive = red, ferromagnetic = yellow, nonmagnetic = blue, individual variance within females of a species = orange, sexual dimorphism = green). Putative magnetoreceptive species are by-definition ferromagnetic, while bees labeled only ferromagnetic have a sigmoidal response but are below the threshold set by *S. quadripunctata*. Non-magnetic bees showed a linear field response. Asterisks (*) indicate species that are nocturnal or crepuscular.

*Anatomical location of magnetic tissues*

For a subset of the female bees, we also evaluated separate body parts for magnetic nanoparticles. The bodies were separated into the metasoma (abdomen), mesosoma (thorax, including wings and legs), and head (including antennae). We evaluated these separate body parts for all four honey bee species (*Apis cerana, A. dorsata, A. florea*, and *A. mellifera*), along with the stingless bee species *Axestotrigona ferruginea* and *Melipona eburnea*, the subsocial *Ceratina strenua*, and the solitary *Centris tamarugales, Euglossa* sp., *Eulaema* sp., and *Ptilothrix bombiformis*. The analysis included the measured variables ($M_S$, $M_S/m$, $H_C$, $M_R$) and also a self-normalized saturation magnetization, representing the distribution of the magnetization within each individual – the percent $M_S$ in each body part: $M_S$(Body Part)/$M_S$(Total). These latter variables correspond to what part of the body the magnetic tissues are located.

These measurements showed that the mesosoma had a significantly higher percent of the saturation magnetization than the head and metasoma, which were not significantly different from one another (Fig. 5A). The head had a significantly lower $H_C$ than the metasoma and mesosoma, which did not differ significantly from one another (Fig. 5B, Table S4). Finally, the head had a lower $M_R/M_S$ compared to the metasoma and the mesosoma, but the mesosoma did not differ from the metasoma (Fig. 5C, Table S4).

Qualitatively, for the 14 individual bees for which we measured separate body parts, we found that in most (7 out of 14) all three body parts were ferromagnetic to some degree (Fig. 5D). The next most likely occurrence (6 out of 14) was that the metasoma and the mesosoma, but not the head, were magnetic. Finally, we observed one occurrence (in *A. cerana*) of the metasoma and head, but not the mesosoma, being magnetic. In our specimens, magnetism was never restricted to a single body part.

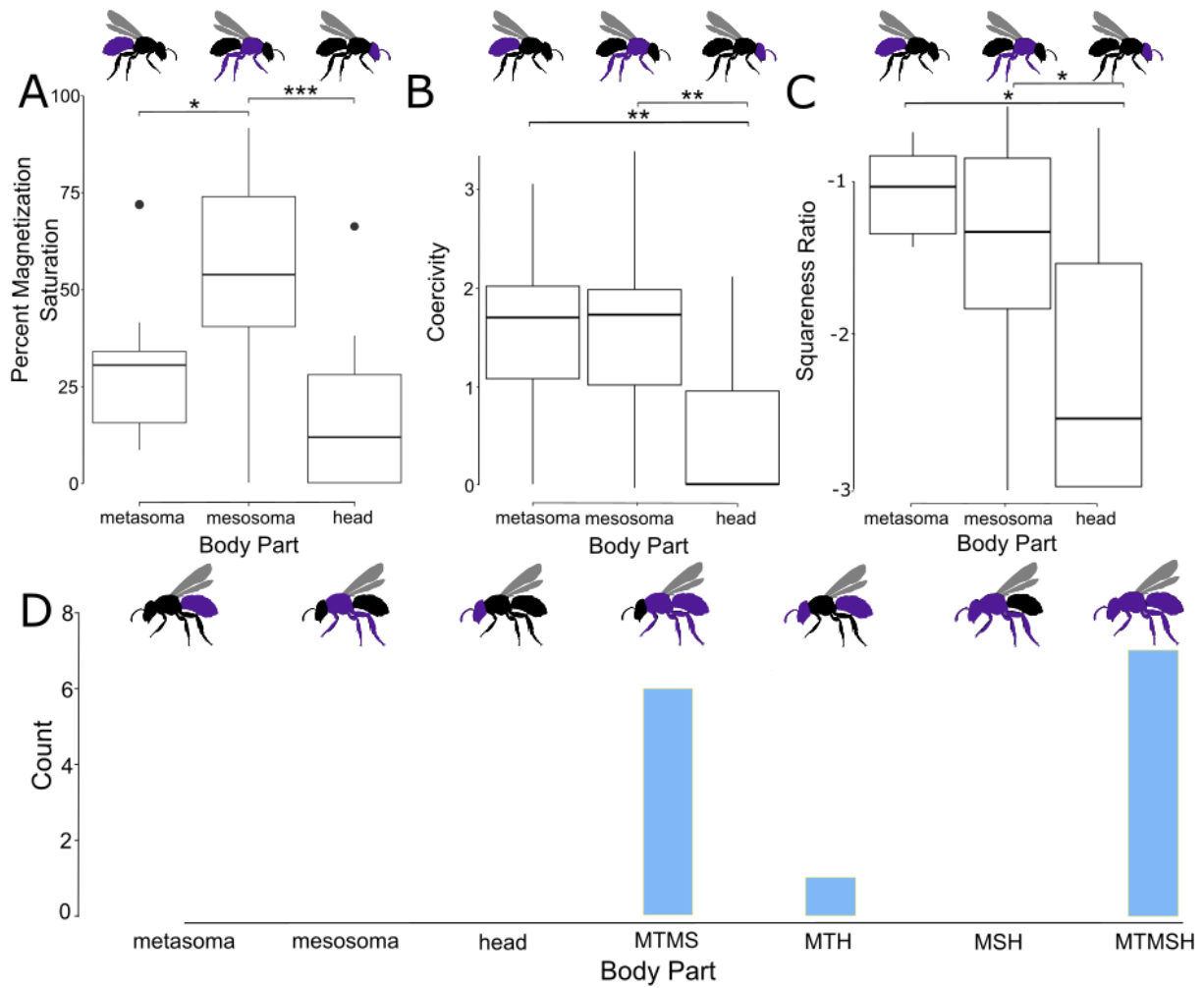

**Figure 5**. **Ferromagnetism is not confined to a single body tagma in bees, but significantly stronger in the mesosoma (thorax w/o wings) and metasoma (abdomen) than the head (w. antennae).** The relative percent of $M_S$ (A), and raw values for $H_C$ (B) and $M_R/M_S$ (C) in the different body tagmata of the bees that were sampled. Significant differences are indicated with asterisks (* < 0.05, ** < 0.01, *** < 0.001). We also include the number of individuals which had ferromagnetic signals in different combinations of body tagmata (MT = metasoma, MS = mesosoma, MTH = metasoma and head, MSH = mesosoma and head, MTMSH = all of the body parts) (D).

The distinct shape of each hysteresis loop for the individual body parts of the specimens (e.g. Fig. 1H) suggested differences in the magnetic nanostructures present in each body region. These differences could reflect the presence of multiple, distinct magnetic tissues within the insect or may arise from environmental iron contamination. However, since environmental contamination would be expected to affect all body parts similarly, it is unlikely that more than one of these signals arose from background sources – supporting the authenticity of the others.

**Discussion**

Our results demonstrate for the first time that magnetism is nearly ubiquitous across bees, and that putative magnetoreception is not restricted to eusocial bee species, or bees from a single taxonomic family. Instead, ferromagnetism was found in different bee species exhibiting a wide range of nesting and social behaviors, as well as cleptoparasitism, and in both males and females. The majority (87.5%) of bee species we assessed exhibited a discernable sigmoidal response indicating the presence of ferromagnetic particles, and most met the threshold for putative magnetoreception, based on values of $M_S$ and $M_S/m$ for *S. quadripunctata* for which magnetoreceptive behavior has been demonstrated (*28*, *56*). Even beyond the bees, in the outgroups we assessed, we detected significant magnetism, including magnetoreceptive potential in solitary (*Cerceris clypeata*, *Bembix sp.*, *Lissonota sp.*, and *Passaloecus sp.*) and eusocial wasps (*Polistes carolina*), along with flies (Syrphidae). However, we also observed several nonmagnetic insects outside of the bee families, including a beetle (*Chauliognathus pensylvanicus*), other social and solitary wasps, and some flies, and overall lower levels of magnetic signal in the non-bee specimens we evaluated. On the other hand, our sampling of these non-bee groups was superficial and does not represent their evolutionary diversity more broadly, especially given the lack of phylogenetic signal, and even individual variance in magnetic signal within our more well-sampled taxa. Other work has demonstrated magnetoreception in some members of these groups; for example, the beetle species *Tenebrio molitor* has been shown to exhibit magnetoreception (*37*).

In our study, the strongest signals of magnetism appeared in the eusocial stingless bees, especially *Melipona fasciata*. However, there were individuals among the stingless bees, *Liotrigona mahaflaya* for example, which were nonmagnetic, which may be a statistical outlier or indicate that stingless bees are not universally magnetic. Notably, every species that was measured in triplicate had at least one ferromagnetic member, indicating that we may have missed a magnetic signal in our "nonmagnetic" species through small sample sizes. Moreover, solitary and even cleptoparasitic bee species were well-represented above our putatively magnetoreceptive threshold, and many were more magnetic than the domesticated western honey bee (*Apis mellifera*). As further demonstration that eusociality and magnetism were not inextricably linked, the eusocial *Augochlorella aurata* was nonmagnetic, while the closely related solitary species *Augochlora pura* was ferromagnetic. Interestingly, there was no phylogenetic signal in the presence or strength of these magnetic signals across the bees. Their widespread pattern suggests that the phylogenetic origin of ferrimagnetic magnetoreception predates the evolution of the Anthophila (bees) (*36*), which likely originated in the Early Cretaceous approximately 125 million years ago (*70*). This situation also leads to questions about the loss of ferrimagnetic magnetoreception in representative bee taxa. For example, we demonstrate cases where congeneric bee species, or even conspecific individuals, are represented by magnetic – and putatively possessing ferrimagnetic magnetoreception – and nonmagnetic specimens (e.g. *Melissodes* *trinodis* and *M. bimaculatus*). There are other mechanisms for magnetoreception in insects, most notably via light-dependent radical pair production in type-I cryptochromes, and a loss of ferrimagnetic magnetoreception while developing another mechanism cannot be ruled out. Yet, bees only have a mammalian-like, light-insensitive type-II cryptochrome (*71*, *72*), which is not consistent with the canonical light-dependent radical pair mechanism.

Natural history traits, such as nesting behavior, sociality, and sex, were therefore not qualitative determinants of the presence or absence of ferromagnetism in bees. However, they did exhibit some influence on the strength of the magnetic signal. For example, the eusocial, cavity-nesting bees tended to exhibit stronger signals of magnetism than solitary, ground-nesting or stem-nesting bees, although solitary bees did have a significantly higher $M_R/M_S$ than eusocial or subsocial/semisocial bees. We also included two nocturnal (*Megaloptidia nocturna* and *Megalopta chaperi*) and one crepuscular (*Eucera pruinosa*) bee species. Although one might hypothesize that night-flying insects might rely more on magnetic fields to navigate (*5*, *73–75*), these bees did not have stronger magnetic signals than other bees, although they met our threshold for putative magnetoreception. We also included measurements of an unusual eusocial, migratory bee (*Apis dosata*). Given the well-established presence of magnetoreception in migratory species, this insect provides a complement to the other bees in the study which were non-migratory. The $M_S$ of *A. dosata* was indeed above the threshold for putative ferrimagnetic magnetoreception, but was not exceptional, and its $M_S/m$ was below our threshold. Interestingly, female worker or solitary bees had significantly higher $H_C$ than male bees, although the three queens we tested did not differ significantly from the workers or males of their species. Given that cleptoparasites do not build or provision their own nests, and therefore do not exhibit central place foraging in the same way that other bees do, it was interesting that they did not differ from solitary and eusocial bee species.

However, cleptoparasites have been shown to return to the same host nest repeatedly, and so may exhibit similar short-range navigational needs (*35*). Perhaps even more interestingly, we observed intraspecific variation in the strength and sometimes presence of the magnetic signal, suggesting magnetism is either a plastic trait that can vary with environmental conditions, such as diet or age (*19*), and/or that populations may vary in the expression of magnetoreception.

Superparamagnetism was also observed among our specimens, although far less commonly than ferromagnetism. Indeed, we observed this only twice, once in a sweat bee (*Augochloropsis metallica*) and once in a specimen of the western honey bee (*A. mellifera*); the other honey bee specimens were all ferromagnetic. Some work has suggested that superparamagetic particles may be less effective than ferromagnetic particles at facilitating magnetoreception (*36*). If superparamagnetism represents some intermediate state between magnetoreceptive and non-magnetoreceptive species, then we would have expected to see a phylogenetic signal. Instead, the presence of individual level variation within a given species (here *A. mellifera*) suggests that this may have more to do with environmental or population-level variance in the development of magnetic particles.

We also observed interesting patterns in the distribution of magnetic nanoparticles among the different body parts of the bees. While other studies have demonstrated that magnetic particles were found in the wings (*53*, *54*), antennae (*30*, *55*, *56*), or metasoma (*30*, *55*), our work showed that, on average, the mesosoma tended to have the highest percentage of magnetic signal among the bees. On the other hand, we never observed a magnetic signal in just one body part. Most commonly, magnetic particles were observed in all three body parts (Fig. 4D), while it was slightly less common for them to be in the metasoma and mesosoma only, and they were never observed in the mesosoma and head only. This contrasts with work in ants, which are closely related to bees and also eusocial, that reported magnetic material across multiple body parts, with the antennae contributing the strongest relative signal in some species (*57–59*). The widespread location of the magnetic nanoparticles, and the relatively lower concentration in the head (which included the light-sensitive organs), may provide support for the presence of a magnetite-mediated mechanism for magnetoreception in the bees (*30*, *52*).

Because our work did not include a behavioral component, we cannot provide definitive proof of magnetoreception beyond establishing the presence of a ferromagnetic signal and comparing it to a threshold based on species for which magnetoreception has been shown (*S. quadripunctata*). This assumption is based on the idea that any organism with a magnetic signal equal to or greater than one with demonstrated magnetoreception is likely to also have the potential to be magnetoreceptive. Anecdotal to this is that there are no other tissues inside insects which are currently known to possess nanoparticles with ferromagnetic properties at room temperature beyond those responsible for magnetoreception. These tissues incur a metabolic cost and so are not expected to be phylogenetically conserved without conferring an advantage. However, it is necessary to test this assumption rigorously with behavioral experiments, although such work is famously difficult (*51*). Importantly, the mere presence of a ferromagnetic signal does not, by itself, imply the existence of magnetoreceptive tissue, as low-level ferromagnetic background from environmental and dietary sources is unavoidable in wild-caught specimens. However, the body part measurements demonstrate that while multiple tagmata may exhibit weak ferromagnetic signals consistent with background contributions, one tagma is typically and significantly more magnetic than the others. If this enriched signal is attributed to magnetoreceptive tissue, then a strong organism-level ferromagnetic signal also reflects the presence of such tissue, with background contributions providing only a minor offset. Accordingly, we use a conservative threshold to distinguish between background-dominated specimens and those likely to possess magnetoreceptive tissues. It is prudent at this point to discuss potential contamination from environmental iron oxide, which could contribute to the observed magnetization. There are several expectations associated with this signal which we did not observe here. First among these is that the data included authentic zeros – in an environment with persistent, widespread ferromagnetic contamination there should be no insects which are authentically nonmagnetic. Next, environmental contamination would be expected on every body part, with a magnetization that loosely follows the surface area of the body part, which again was not the case. Third, ground-nesting bees would be expected to have higher magnetization due to their constant exposure to iron oxide in the soil and from environmental pollution, and this was not the case. Fourth, mass scales with the volume of the bee, while the potential for surface contamination would scale with its surface area, a slower growing mathematical function. Thus, the mass-specific magnetization, $M_S/m$, should be a decreasing function for surface contamination, while our data showed it significantly ($p < 0.05$) increasing. Finally, the body part measurements showed different magnetic behavior in each body part. If there were environmental contamination, we would expect a similar signal between each body part. As a final thought, known magnetoreceptive insects, such as *B. impatiens* (*25*) and *A. mellifera*, would have a magnetic signal equal to that of the background plus the magnetoreceptive tissues. The fact that these insects were not exceptional in the range of magnetization distribution, Fig. 3, implies that the other magnetic species also had contributions from both. Put another way, insects with only background sources would be expected to have a much lower signal than *B. impatiens* (*25*) and *A. mellifera*, but most bees did not.

As noted above, ferromagnetic nanoparticles are not the only proposed mechanism to achieve magnetoreception. In other organisms, such as *Drosophila*, cryptochromes have been found in photoreceptive parts of the insect, including the eyes and antennae (*71*). Importantly, the presence and anatomical distribution of ferromagnetic signals measured here do not exclude the possibility of radical-pair based magnetoreception operating independently in bees. Rather, our results constrain the origin of the measured magnetic signal, indicating the presence of ferrimagnetic nanoparticles in non-photoreceptive body regions such as the metasoma and mesosoma. The detection of these nanoparticles is therefore consistent with magnetite-based models of magnetoreception, although we cannot determine from these measurements alone whether such particles play a sensory role, a supporting role, or coexist with other magnetoreceptive mechanisms. Some work on honey bees suggested that trophocyte cells in the abdominal cuticle involved in the storage of excess dietary iron might relate to magnetic particle-mediated magnetoreception (*48*) and this may also be a factor in magnetoreception in other bees.

Broadly, our work demonstrates that magnetic properties, and very likely magnetoreception, are a common feature, at least among the bees, but also probably in wasps and other groups of insects, as indicated by our outgroups. In bees, magnetism and putative magnetoreception do not appear to be determined by a particular life history or social behavior, and while the strength of these magnetic signals may have averaged slightly higher in eusocial species, they were found across a broad range of taxa, including both solitary and cleptoparasitic species. Future work may explore the fitness implications for these magnetic nanoparticles, their ability to influence a bee's foraging and dispersal capability, and whether or not they can be affected by diet or some other unknown environmental attribute.

*Acknowledgments:*

We would like to acknowledge assistance from S. Droege (USGS) in identifying bee specimens collected by LR. We thank those who assisted with insect collections, including D. Eldridge, A. Lawson, K. McKim, N. Oldham, A. Khalil, E. Dalliance, S. Wilhelm, and S. Baldwin. We also would like to thank the sites where the specimens were collected, including the UT Research and Education Centers (Plateau, Organic Crops Unit, Forest Resources), the UT Gardens, and the Great Smoky Mountains Institute at Tremont. MW acknowledges funding by the DFG (SFB 1372, project no. 395940762)

*Author Contributions:*

Conceptualization: DAG, AFM, LR, LP, SB, MW

Methodology: LR, LP, SB, MW, AFM, DAG

Investigation: DAG, AFM, LR, LP, SB, MW

Visualization: LR, DAG

Funding acquisition: LR, DAG

Project administration: LR, DAG, AFM

Supervision: LR, DAG, AFM

Writing – original draft: LR, DAG

Writing – review & editing: DAG, AFM, LR, LP, SB, MW

*Competing interests:*

The authors have no competing interests to declare

*Data and materials availability:*

Data are available at FigShare at: https://figshare.com/s/132e66beb487b4c7f40a